\documentclass[sigconf]{acmart}
\AtBeginDocument{%
  }

\setcopyright{none}
\copyrightyear{2018}
\acmYear{2018}
\acmDOI{XXXXXXX.XXXXXXX}
\acmConference[Conference acronym 'XX]{Make sure to enter the correct
  conference title from your rights confirmation email}{June 03--05,
  2018}{Woodstock, NY}
\acmISBN{978-1-4503-XXXX-X/2018/06}
\usepackage{algorithm}
\usepackage{algorithmic}
\usepackage{booktabs}
\usepackage{longtable}
\usepackage{bm}
\usepackage{xcolor}
\usepackage{multirow}
\usepackage{graphicx}
\usepackage{makecell}
\usepackage{subcaption}
\usepackage{lipsum}
\usepackage{enumitem}
\usepackage{amsmath}
\usepackage{amsfonts}
\usepackage{threeparttable}

\usepackage{amssymb}

\begin{document}

\title{Intent-Guided Reasoning for Sequential Recommendation}

\author{Yifan Shao}
\affiliation{%
  \institution{The Chinese University of Hong Kong}
  \city{Hong Kong}
  \country{China}
  }

\email{yifashao209@gmail.com}

\author{Peilin Zhou}

\affiliation{%
  \institution{Hong Kong University of Science and Technology (Guangzhou)}
  \city{Guangzhou}
  \country{China}
}
\email{zhoupalin@gmail.com}

\renewcommand{\shortauthors}{Shao et al.}

\begin{abstract}
Sequential recommendation systems aim to capture users' evolving preferences from their interaction histories. Recent reasoning-enhanced methods have shown promise by introducing deliberate, chain-of-thought-like processes with intermediate reasoning steps. However, these methods rely solely on the next target item as supervision, leading to two critical issues: (1) \textit{reasoning instability}—the process becomes overly sensitive to recent behaviors and spurious interactions like accidental clicks, and (2) \textit{surface-level reasoning}—the model memorizes item-to-item transitions rather than understanding intrinsic behavior patterns.
To address these challenges, we propose \textbf{IGR-SR}, an \textbf{I}ntent-\textbf{G}uided \textbf{R}easoning framework for \textbf{S}equential \textbf{R}ecommendation that anchors the reasoning process to explicitly extracted high-level intents. Our framework comprises three key components: (1) a Latent Intent Distiller (LID) that efficiently extracts multi-faceted intents using a frozen encoder with learnable tokens, (2) an Intent-aware Deliberative Reasoner (IDR) that decouples reasoning into intent deliberation and decision-making via a dual-attention architecture, and (3) an Intent Consistency Regularization (ICR) that ensures robustness by enforcing consistent representations across different intent views. Extensive experiments on three public datasets demonstrate that IGR-SR achieves an average 7.13\% improvement over state-of-the-art baselines. Critically, under 20\% behavioral noise, IGR-SR degrades only 10.4\% compared to 16.2\% and 18.6\% for competing methods, validating the effectiveness and robustness of intent-guided reasoning.
\end{abstract}


\maketitle

\section{Introduction}
\label{sec:introduction}

Reasoning-enhanced sequential recommendation methods~\cite{tang2025thinkrecommendunleashinglatent, liu2025lareslatentreasoningsequential} have recently emerged as a promising paradigm. Unlike the traditional reflexive paradigm (\textit{e.g.}, SASRec~\cite{kang2018self}, BERT4Rec~\cite{sun2019bert4rec}), which directly maps historical behaviors to a unified prediction, these new models simulate a deliberate, chain-of-thought-like process. By generating intermediate reasoning steps, they model a user's latent decision path, showing significant potential for capturing complex, evolving user preferences.

Despite their potential, existing reasoning-enhanced methods suffer from a fundamental limitation: their reasoning process is guided solely by the next target item. This reliance on a single, often noisy, supervisory signal creates two fundamental problems. First, it leads to \textbf{reasoning instability}, making the model overly sensitive to recent interactions or spurious clicks and causing it to drift from the user's true long-term goals. Second, it encourages \textbf{surface-level reasoning}, where the model simply memorizes item-to-item transitions (\textit{e.g.}, ``after item A comes item B'') instead of understanding intrinsic behavior patterns behind the sequence (\textit{e.g.}, ``exploring outdoor activities''). This naturally raises a crucial question: \textit{How can we guide the reasoning process with a more stable, high-level signal?}

We argue that the key lies in extracting and leveraging abstract user intents---the underlying goals driving a user's behavior. Ideally, anchoring the reasoning process to these stable intents can filter out the noise from short-term actions and promote the learning of generalizable preference patterns. However, putting this idea into practice is non-trivial, as it presents two major technical hurdles: (1) How to efficiently extract user intents, without introducing significant computational and parametric overhead? (2) How to effectively integrate this high-level guidance into a step-by-step reasoning process without disrupting its sequential nature?

To address these challenges, we propose \textbf{IGR-SR}, a novel and effective \textbf{I}ntent-\textbf{G}uided \textbf{R}easoning framework for \textbf{S}equential \textbf{R}ecom\\mendation. For efficient intent extraction, we design a \textbf{Latent Intent Distiller (LID)}. This module adds a small set of learnable prefixes to steer a frozen pre-trained encoder, guiding it to distill multi-faceted intents into dedicated \texttt{<intent>} tokens at minimal cost. For effective fusion, we propose an \textbf{Intent-aware Deliberative Reasoner (IDR)}, which explicitly decouples reasoning into two synergistic stages: \textit{intent deliberation} and \textit{decision-making}. This is realized through a dual-attention architecture: cross-attention enables deliberation by dynamically integrating relevant global intents, then masked self-attention handles decision-making by capturing local sequential patterns. Finally, to enhance the robustness of  intent guidance, we further introduce an \textbf{Intent Consistency Regularization (ICR)} objective. This regularization technique ensures the final user representation remains consistent across different masked views of the user's intents, preventing over-reliance on specific intent subsets. Our contributions are threefold:
\begin{itemize}
\item We propose \textbf{IGR-SR}, a novel intent-guided Reasoning framework for sequential recommendation that addresses the critical limitation of unguided reasoning by explicitly anchoring the deliberative reasoning process to high-level user intents.
\item We design a Latent Intent Distiller (LID) for efficient intent distillation and an Intent-aware Deliberative Reasoner (IDR) for effective, intent-guided reasoning. The framework's robustness is further enhanced by an Intent Consistency Regularization (ICR) objective.
\item Extensive experiments on three public datasets demonstrate that our approach consistently outperforms state-of-the-art baselines and shows superior robustness against behavioral noise.
\end{itemize}

\section{Problem Formulation}
Let $\mathcal{U}$ be the set of users and $\mathcal{I}$ be the set of items. For each user $u \in \mathcal{U}$, we are given their historical interaction sequence $S^u = [i_1, i_2, \dots, i_n]$ in chronological order, where $n$ is the sequence length. The objective of sequential recommendation is to predict the next item $i_{n+1}$ that the user is most likely to interact with.

\section{Methodology}
\label{sec:methodology}

\begin{figure}[t]
    \centering
    \includegraphics[width=0.5\textwidth]{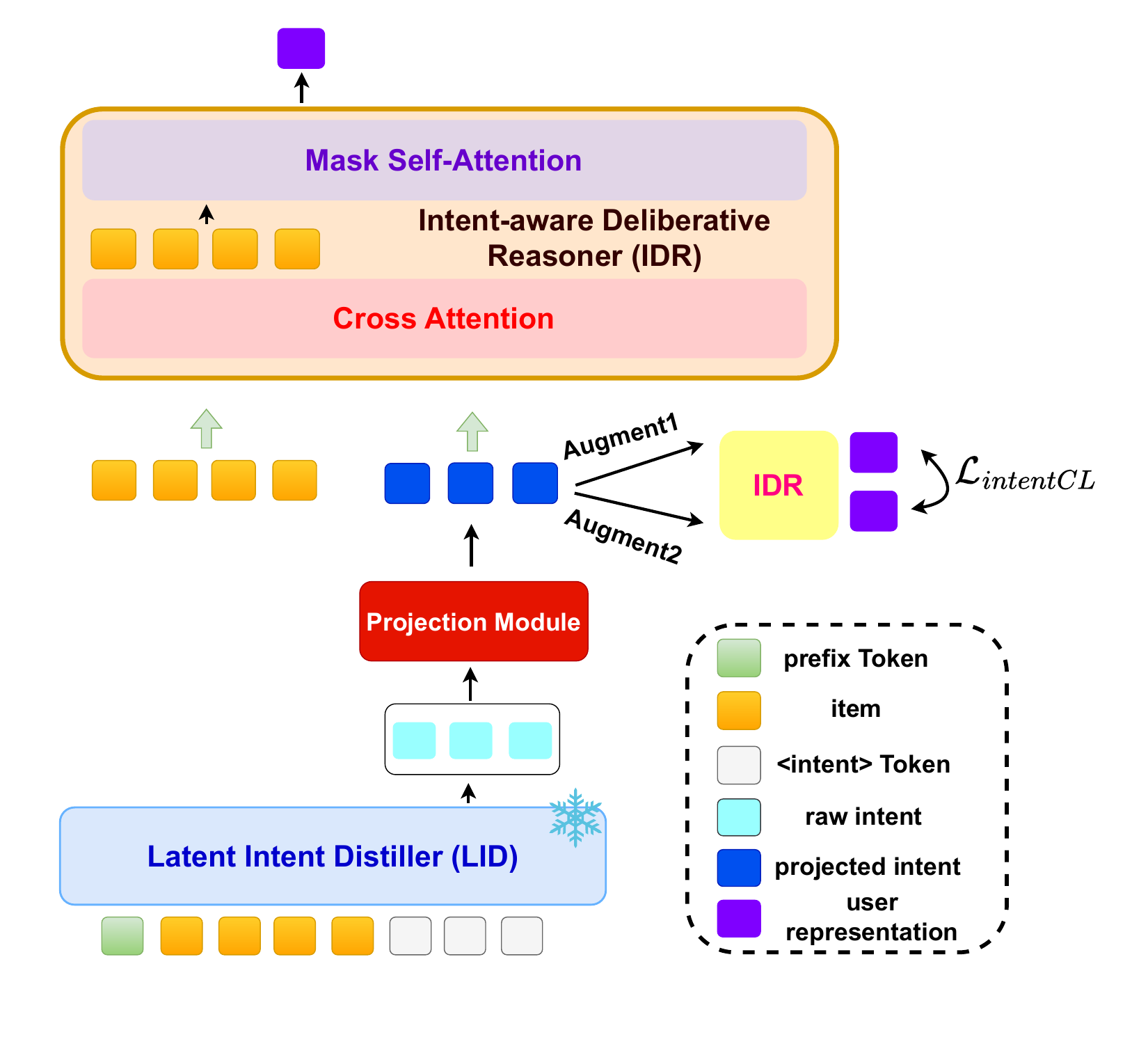}
    \vspace{-0.3in}
    \caption{The overall architecture of our IGR-SR framework.}
    \label{figure:overview}
\end{figure}

Our IGR-SR framework comprises three core components: a Latent Intent Distiller (LID), an Intent-aware Deliberative Reasoner (IDR) and an Intent Consistency Regularization (ICR) objective. The LID first distills user intents from the raw sequence, which are then fused by the IDR to perform intent-guided reasoning. Finally, the ICR is introduced to ensure the robustness of intent guidance. The overall architecture of IGR-SR is illustrated in Figure~\ref{figure:overview}.

\subsection{Latent Intent Distiller (LID)}
Inspired by prompt tuning and SoftCot~\cite{xu2025softcot} in LLMs, we designed the Latent Intent Distiller (LID) for the lightweight and efficient extraction of user intents. Specifically, we first add $k$ learnable prefix tokens to the beginning of the item sequence, denoted as $\mathcal{P} = [p_1, \dots, p_k]$. Their corresponding embeddings are optimized to serve as task-specific instructions, steering the frozen, pre-trained SASRec encoder towards distilling latent user intents. Then, we append $m$ special \texttt{<intent>} tokens to the end of the sequence. These tokens serve as dedicated placeholders designed to aggregate and embody the extracted intent information.

Formally, we first construct the augmented input sequence $S^u_{\text{aug}}$ as follows:
\begin{equation}
    \label{eq:seq_concat}
    S^u_{\text{aug}} = \text{Concat}(\mathcal{P}, S^u, \mathcal{I}),
\end{equation}
where $\mathcal{P}$ is the sequence of $k$ prefix tokens, $S^u$ is the user's item sequence, and $\mathcal{I}$ is a sequence of $m$ \texttt{<intent>} tokens. $S^u_{\text{aug}}$ is then processed by the LID Encoder, which internally handles embedding lookup and positional encoding, producing the final hidden states $\mathbf{H}_{\text{I}} \in \mathbb{R}^{(k+n+m) \times d_{\text{I}}}$ :
\begin{equation}
    \label{eq:ien_encoder_simple}
    \mathbf{H}_{\text{I}} = \text{LID}(S^u_{\text{aug}}),
\end{equation}
The intent representations $ \mathbf{T}_{\text{I}}$ are the output hidden states at the positions corresponding to the \texttt{<intent>} tokens:
\begin{equation}
    \label{eq:ien_output}
    \mathbf{T}_{\text{I}} = {\mathbf{H}_{\text{I}}}_{[k+n+1 : k+n+m]} \in \mathbb{R}^{m \times d_{\text{I}}}.
\end{equation}
Critically, the parameters of the LID Encoder remain frozen. Gradients from the loss flow back to optimize only the embeddings of prefix and \texttt{<intent>} tokens. This design makes the LID both highly effective and computationally efficient.

\subsection{Projection Module}
Since there exists both a representation gap and a dimensional gap between the LID and the main model (IDR), a direct utilization of $\mathbf{T}_{\text{I}}$ may lead to suboptimal performance. To bridge this gap, we introduce a projection module that maps these raw intents to the representation space of the main network.

\begin{equation}
    \mathbf{T}_{\text{D}} = f_\theta(\mathbf{T}_{\text{I}}) \in \mathbb{R}^{m \times d}.
\end{equation}
where $d$ is the hidden dimension of the main network and $f_\theta: \mathbb{R}^{d_{\text{I}}} \rightarrow \mathbb{R}^{d}$ is implemented as a lightweight Multi-Layer Perceptron (MLP). The projected intents $\mathbf{T}_{\text{D}}$ serve as the stable, high-level ``knowledge base'' for the subsequent decision-making process.

\subsection{Intent-aware Deliberative Reasoner (IDR)}

The IDR serves as the main recommendation model in our framework, responsible for performing the intent-guided reasoning. To achieve effective intent fusion, the proposed IDR explicitly decouples the recommendation process into two independent yet synergistic stages: \textit{Intent Deliberation} and \textit{Decision-Making}. This separation is realized through a novel dual-attention architecture. The IDR takes the original user item sequence $S^u$ as its input. The initial representations $\mathbf{H}_{\text{D}}^{(0)}$ are then processed through a stack of $L$ dual-attention layers.

\subsubsection{Intent Deliberation}
The first stage is responsible for enriching the local context with global user intents. Instead of the simple concatenation of the item sequence and extracted intents, which may suffer from positional encoding contamination, we employ a cross-attention mechanism to inject the intent information. Here, the sequence representations $\mathbf{H}_{\text{D}}^{(l-1)}$ act as \textit{queries}, while the projected intent vectors $\mathbf{T}_{\text{D}}$ serve as both \textit{keys} and \textit{values}. This allows each item's representation to dynamically retrieve and integratet he most relevant intent information for its current context:
\begin{equation}
    \mathbf{H}_{\text{cross}}^{(l)} = \mathbf{H}_{\text{D}}^{(l-1)} + \text{CrossAttn}(Q=\mathbf{H}_{\text{D}}^{(l-1)}, K=\mathbf{T}_{\text{D}}, V=\mathbf{T}_{\text{D}}),
\end{equation}

\subsubsection{Decision-Making}
Following the intent deliberation, the second stage focuses on capturing the sequential dynamics within the now intent-enriched representations. A standard masked self-attention mechanism is applied to model the temporal dependencies:
\begin{equation}
    \mathbf{H}_{\text{self}}^{(l)} = \mathbf{H}_{\text{cross}}^{(l)}+\text{MaskedSelfAttn}(\mathbf{H}_{\text{cross}}^{(l)}),
\end{equation}
The output then passes through a feed-forward network (FFN) with layer normalization to produce the output of the layer:
\begin{equation}
    \mathbf{H}_{\text{D}}^{(l)} = \text{LayerNorm}\left(\mathbf{H}_{\text{self}}^{(l)} + \text{FFN}(\mathbf{H}_{\text{self}}^{(l)})\right).
\end{equation}
After processing through all $L$ layers, we take the final hidden state of the last item (\textit{i.e.}, $ \mathbf{H}_{\text{D}}^{(L)}[n]$), as the user's ultimate representation $\mathbf{h}_u$.

\subsection{Intent Consistency Regularization (ICR)}
To enhance the robustness of intent-guided deliberation, we introduce an \textbf{Intent Consistency Regularization (ICR)} mechanism. Specifically, we employ a contrastive learning objective that forces the model to produce consistent user representations from different subsets of intent information, preventing over-reliance on specific intents.

\subsubsection{Masked Intent Augmentation}
Given the projected intents $\mathbf{T}_{\text{D}}$, we generate two augmented views by applying independent random masks:

\begin{equation}
    \mathbf{T}_{\text{D}}^{(1)} = \mathbf{T}_{\text{D}} \odot \mathbf{M}^{(1)}, \quad
    \mathbf{T}_{\text{D}}^{(2)} = \mathbf{T}_{\text{D}} \odot \mathbf{M}^{(2)},
\end{equation}
where $\mathbf{M}^{(i)} \in \{0,1\}^{m \times d}$ are independently sampled masks with dropout probability $p_{\text{mask}}$. Each masked intent set is fed into the IDR to obtain user representations:

\begin{equation}
    \mathbf{h}_{u}^{(1)} = \text{IDR}(S^u; \mathbf{T}_{\text{D}}^{(1)}), \quad
    \mathbf{h}_{u}^{(2)} = \text{IDR}(S^u; \mathbf{T}_{\text{D}}^{(2)}),
\end{equation}

\subsubsection{Contrastive Objective}
We maximize agreement between the two views using InfoNCE loss:
\begin{equation}
    \mathcal{L}_{\text{IntentCL}} = -\sum_{u \in \mathcal{U}} \log \frac{\exp(\text{sim}(\mathbf{h}_{u}^{(1)}, \mathbf{h}_{u}^{(2)}) / \tau)}{\sum_{v \in \mathcal{U}} \exp(\text{sim}(\mathbf{h}_{u}^{(1)}, \mathbf{h}_{v}^{(2)}) / \tau)}.
\end{equation}
where $\text{sim}(\cdot, \cdot)$ denotes the cosine similarity, and $\tau$ is a temperature hyperparameter. The final objective combines standard recommendation and contrastive losses.

\section{Experiments}
\subsection{Experimental Setup}
\subsubsection{Datasets and Evaluation Metrics}
We conduct experiments on three real-world recommendation datasets: Toys, Instrument and CDs\_and\_Vinyl, which are constructed from Amazon review datasets. The detailed statistics of the datasets are summarized in Table~\ref{table:dataset}. For a fair comparison, we follow the same data preprocessing and evaluation metrics as in previous work~\cite{tang2025thinkrecommendunleashinglatent}.

\subsubsection{Compared Models}
We compare IGR-SR with several competitive baseline models. \textbf{Basic Models} including GRU4Rec~\cite{hidasi2015session}, BERT4Rec~\cite{sun2019bert4rec}, and SASRec~\cite{kang2018self}, which model user sequences using recurrent or Transformer-based architectures; \textbf{Intent-based Models}: ICLRec~\cite{chen2022intent} and ICSRec~\cite{qin2024intent}, which explicitly model user intent in non-reasoning framework and \textbf{Reasoning-enhanced Models}: ReaRec~\cite{tang2025thinkrecommendunleashinglatent} and LARES~\cite{liu2025lareslatentreasoningsequential}, which introduce intermediate reasoning steps to simulate deliberate decision-making.

\subsubsection{Implementation Details}
For a fair comparison, the hyperparameter settings for all baselines are adopted from their original papers. For our LID module, we use a frozen pre-trained SASRec encoder with dimension $d_{\text{I}}$ chosen from \{8, 16, 32\}. The number of prefix tokens $k$ and \texttt{<intent>} tokens $m$ are tuned in $[2, 32]$ and $[1,5]$, respectively. The hidden dimension of IDR is set as $d=64$. For ICR, we tune the $p_{\text{mask}}$ in $[0.1,0.8]$.

\begin{table}[!t]
    \caption{Statistics of the datasets}
    \vspace{-0.15in}
	\begin{tabular}{c *{4}{r}}
		\toprule
		\textbf{Datasets} & \textbf{\#Users} & \textbf{\#Items} & \textbf{\#Interactions} & \textbf{Sparsity}\\
		\midrule
		Toys 	& 19,412  & 11,925  & 167,597   & 99.93\%  \\
		Instrument 	& 57,439 & 24,587 & 511,836 & 99.96\% \\
		CDs\_and\_Vinyl & 75,258 & 64,443 & 1097,592 & 99.98\% \\
		\bottomrule
	\end{tabular}
    \label{table:dataset}
\end{table}

\begin{table*}[t]
\captionsetup{font={small}}
\caption{Performance comparison of different methods. The best and second-best results are indicated in bold and underlined font, respectively. ``*'' denotes that the improvements are statistically significant with $p < 0.01$ in a paired t-test setting.}
\vspace{-0.1in}
\resizebox{0.9\textwidth}{!}{%
  \renewcommand\arraystretch{0.98}  
  \setlength{\tabcolsep}{2mm}       
  \begin{tabular}{llcccccccccc}
    \toprule
    Dataset & Metric & GRU4Rec & BERT4Rec & SASRec & ICLRec & ICSRec & ReaRec & LARES & IGR-SR \\ 
    \midrule

    \multirow{4}{*}{Toys} 
      & Recall@10  & 0.0449 & 0.0314 & 0.0708 & 0.0716 & 0.0711 & 0.0723 & \underline{0.0731} & \textbf{0.0802*} \\
      & Recall@20  & 0.0722 & 0.0493 & 0.1022 & 0.1027 & 0.1024 & 0.1042 & \underline{0.1046} & \textbf{0.1149*} \\
      & NDCG@10   & 0.0223 & 0.016 & 0.0344 & 0.0348 & 0.0342 & 0.0351 & \underline{0.0354} & \textbf{0.0372*} \\
      & NDCG@20   & 0.0291 & 0.0205 & 0.0423 & 0.0428 & 0.0426 & 0.0426 & \underline{0.0432} & \textbf{0.0460*} \\
    \midrule

    \multirow{4}{*}{Instrument} 
      & Recall@10  & 0.0498 & 0.04 & 0.0517 & 0.0528 & 0.0521 & \underline{0.0531} & 0.0523 & \textbf{0.0562*} \\
      & Recall@20  & 0.0751 & 0.0614 & 0.0758 & 0.0767 & 0.0762 & \underline{0.0774} & 0.0770 & \textbf{0.0828*} \\
      & NDCG@10   & 0.0259 & 0.0209 & 0.0267 & 0.0274 & 0.0269 & \underline{0.0277} & 0.0271 & \textbf{0.0289*} \\
      & NDCG@20   & 0.0323 & 0.0263 & 0.0328 & 0.0336 & 0.0331 & \underline{0.0341} & 0.0334 & \textbf{0.0362*} \\
    \midrule

    \multirow{4}{*}{CDs\_and\_Vinyl} 
      & Recall@10  & 0.0608 & 0.0481 & 0.0855 & 0.0871 & 0.0872 & 0.0852 & \underline{0.0874} & \textbf{0.0921*} \\
      & Recall@20  & 0.0945 & 0.0719 & 0.1290 & 0.1302 & 0.1296 & 0.1286 & \underline{0.1317} & \textbf{0.1388*} \\
      & NDCG@10   & 0.0307 & 0.0248 & 0.0383 & 0.0388 & 0.0389 & 0.0384 & \underline{0.0392} & \textbf{0.0406*} \\
      & NDCG@20   & 0.0392 & 0.0305 & 0.0490 & 0.0494 & 0.0497 & 0.0486 & \underline{0.0501} & \textbf{0.0525*} \\
    \bottomrule
  \end{tabular}
}

\label{table:overall}
\end{table*}

\subsection{Overall Performance}
Table~\ref{table:overall} reports the performance of all compared methods across three datasets. Key observations are summarized as follows:

\begin{itemize}
    \item Intent-based methods like ICLRec and ICSRec show consistent improvements over base models. This demonstrates that by capturing this higher-level abstraction, the models can better understand user behavior.
    
    \item Reasoning-enhanced approaches ReaRec and LARES achieve notable improvement by simulating a deliberate reasoning process before the final prediction, validating that this think-before-action paradigm in LLMs is equally effective in sequential recommendation.
    
    \item Our proposed IGR-SR consistently outperforms all baselines by explicitly guiding the reasoning process with latent user intents. Anchoring deliberation reasoning with high-level intent information significantly enhances recommendation performance across datasets.
\end{itemize}

\begin{figure}[t]
  \centering
  \includegraphics[width=\linewidth]{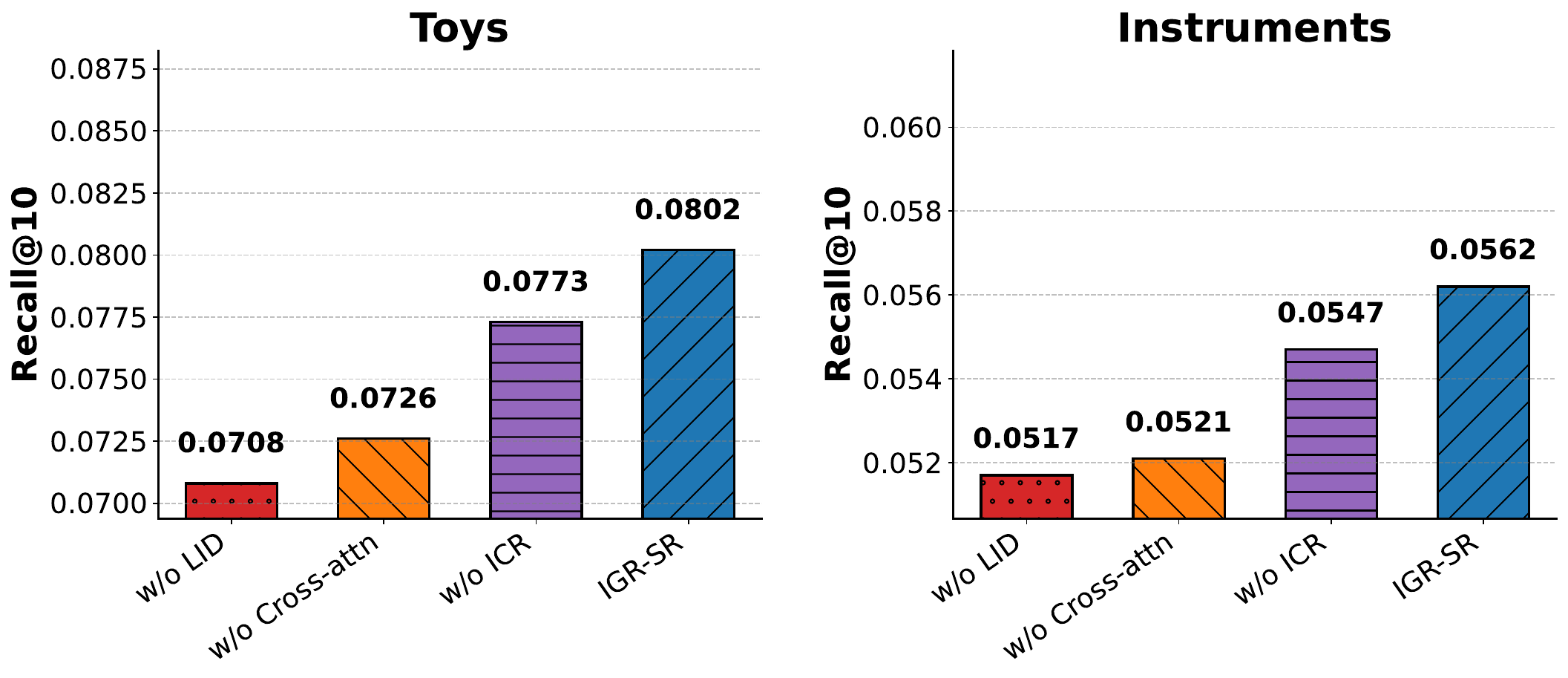}
  \vspace{-0.3in}
  \caption{Ablation study for different components of IGR-SR.}
  \label{figure:ablation}
\end{figure}

\begin{figure}[h]
  \centering
  \includegraphics[width=\linewidth]{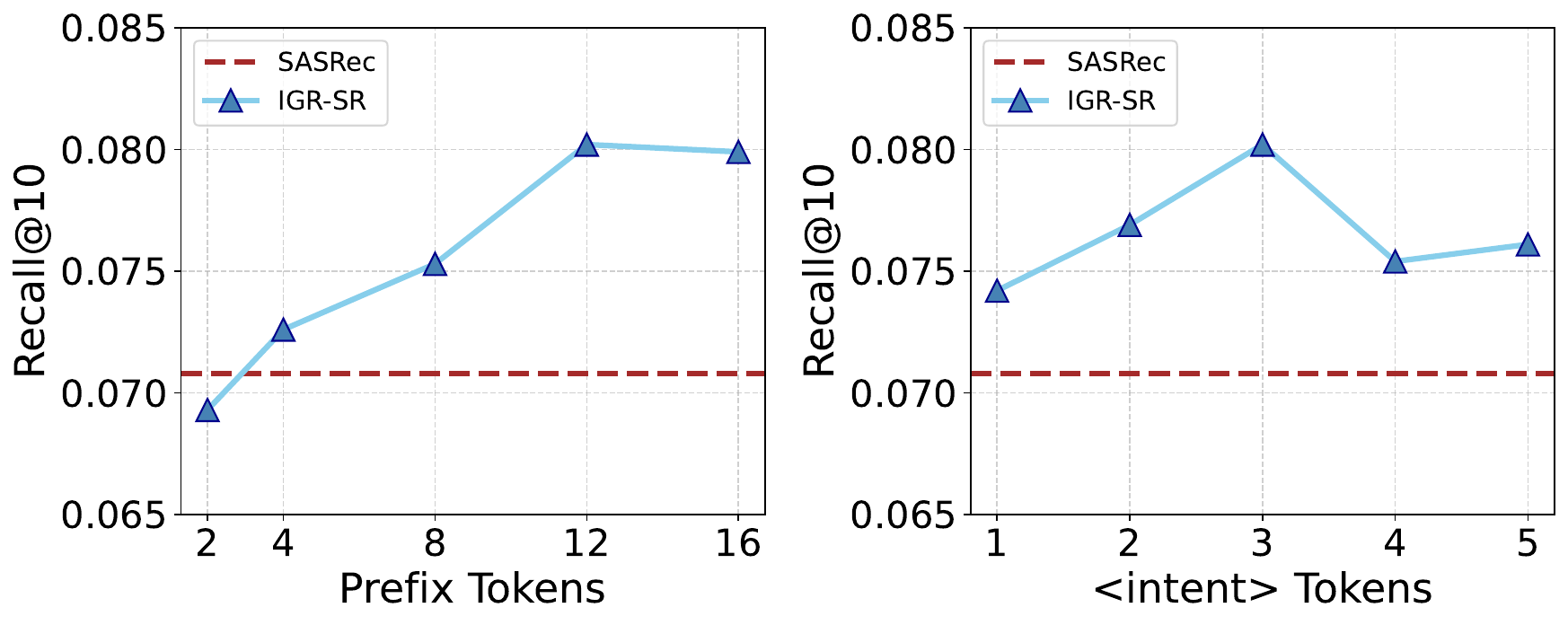}
  \vspace{-0.2in}
  \caption{Effect of varying prefix and \texttt{<intent>} token numbers on Recall@10.}
  \label{figure:hyperparameter}
\end{figure}

\subsection{Ablation Study}
Figure~\ref{figure:ablation} presents the ablation study on IGR-SR. Each variant removes one component: ``w/o LID'' removes the Latent Intent Distiller, ``w/o cross-attn'' replaces the cross-attention with direct concatenation of the item sequence and extracted intents, and ``w/o ICR'' disables the intent-mask contrastive learning. The results show that removing any component leads to a performance decrease, indicating their contributions.

\subsection{Further Analysis}
\subsubsection{Noise Robustness}
To assess the model's resilience, we simulated noisy user behavior by randomly perturbing 20\% of sequential interactions. As shown in Table~\ref{table:noise}, while all models experience performance degradation, IGR-SR exhibits a significantly smaller relative drop compared to SASRec and ReaRec. This result validates that by anchoring the reasoning process to stable, high-level intents, our model can effectively mitigate the impact of spurious actions.

\subsubsection{Impact of the Number of Prefix and \texttt{<intent>} Tokens}
We investigated the influence of the number of prefix tokens ($k$) and \texttt{<intent>} tokens ($m$). As shown in Figure~\ref{figure:hyperparameter}, performance is suboptimal with too few tokens. A small $k$ fails to provide effective signals to guide the frozen encoder for intent distillation, while a small $m$ lacks the capacity to capture diverse user intents. 

\section{Conclusion}
In this paper, we introduced IGR-SR, a novel and effective framework that addresses the critical issue of unguided reasoning in sequential recommendation. By explicitly distilling multi-faceted user intents and using them to guide a deliberative reasoning process, IGR-SR achieves more robust and accurate recommendation predictions, proving particularly resilient against noisy interactions and short-term behavioral patterns.

\begin{table}[t]
\centering
\small
\caption{Robustness Analysis under 20\% noise (Recall@10).}
\vspace{-0.1in}
\begin{tabular}{lcccc}
\toprule
\multirow{2}{*}{Method} & \multicolumn{2}{c}{Toys} & \multicolumn{2}{c}{Instrument} \\
\cmidrule(lr){2-3} \cmidrule(lr){4-5}
 & Clean & +20\% Noise & Clean & +20\% Noise \\
\midrule
SASRec & 0.0708 & 0.0576 (↓18.6\%) & 0.0517 & 0.0454 (↓12.2\%) \\
ReaRec & 0.0723 & 0.0606 (↓16.2\%) & 0.0531 & 0.0468 (↓11.9\%) \\
IGR-SR   & \textbf{0.0802} & \textbf{0.0719 (↓10.4\%)} & \textbf{0.0562} & \textbf{0.0513 (↓8.7\%)} \\
\bottomrule
\end{tabular}
\label{table:noise}
\end{table}

\bibliographystyle{ACM-Reference-Format}
\bibliography{ref}

@inproceedings{kang2018self,
  title={Self-attentive sequential recommendation},
  author={Kang, Wang-Cheng and McAuley, Julian},
  booktitle={2018 IEEE international conference on data mining (ICDM)},
  pages={197--206},
  year={2018},
  organization={IEEE}
}

@inproceedings{sun2019bert4rec,
  title={BERT4Rec: Sequential recommendation with bidirectional encoder representations from transformer},
  author={Sun, Fei and Liu, Jun and Wu, Jian and Pei, Changhua and Lin, Xiao and Ou, Wenwu and Jiang, Peng},
  booktitle={Proceedings of the 28th ACM international conference on information and knowledge management},
  pages={1441--1450},
  year={2019}
}

@article{hidasi2015session,
  title={Session-based recommendations with recurrent neural networks},
  author={Hidasi, Bal{\'a}zs and Karatzoglou, Alexandros and Baltrunas, Linas and Tikk, Domonkos},
  journal={arXiv preprint arXiv:1511.06939},
  year={2015}
}

@inproceedings{chen2022intent,
  title={Intent contrastive learning for sequential recommendation},
  author={Chen, Yongjun and Liu, Zhiwei and Li, Jia and McAuley, Julian and Xiong, Caiming},
  booktitle={Proceedings of the ACM web conference 2022},
  pages={2172--2182},
  year={2022}
}

@misc{tang2025thinkrecommendunleashinglatent,
      title={Think Before Recommend: Unleashing the Latent Reasoning Power for Sequential Recommendation}, 
      author={Jiakai Tang and Sunhao Dai and Teng Shi and Jun Xu and Xu Chen and Wen Chen and Wu Jian and Yuning Jiang},
      year={2025},
      eprint={2503.22675},
      archivePrefix={arXiv},
      primaryClass={cs.IR},
      url={https://arxiv.org/abs/2503.22675}, 
}

@misc{liu2025lareslatentreasoningsequential,
      title={LARES: Latent Reasoning for Sequential Recommendation}, 
      author={Enze Liu and Bowen Zheng and Xiaolei Wang and Wayne Xin Zhao and Jinpeng Wang and Sheng Chen and Ji-Rong Wen},
      year={2025},
      eprint={2505.16865},
      archivePrefix={arXiv},
      primaryClass={cs.IR},
      url={https://arxiv.org/abs/2505.16865}, 
}

@article{xu2025softcot,
  title={Softcot: Soft chain-of-thought for efficient reasoning with llms},
  author={Xu, Yige and Guo, Xu and Zeng, Zhiwei and Miao, Chunyan},
  journal={arXiv preprint arXiv:2502.12134},
  year={2025}
}

@inproceedings{qin2024intent,
  title={Intent contrastive learning with cross subsequences for sequential recommendation},
  author={Qin, Xiuyuan and Yuan, Huanhuan and Zhao, Pengpeng and Liu, Guanfeng and Zhuang, Fuzhen and Sheng, Victor S},
  booktitle={Proceedings of the 17th ACM international conference on web search and data mining},
  pages={548--556},
  year={2024}
}
\end{document}